\documentclass{llncs}
\usepackage{url}
\usepackage{latexsym,amsmath,amssymb,stmaryrd}

\usepackage{xspace}
\usepackage{enumitem}
\usepackage[caption=false]{subfig}
\usepackage{listings}
\usepackage{pgf}
\usepackage{mathpartir}
\usepackage{color}
\usepackage{soul}
\usepackage{subscript}
\usepackage[normalem]{ulem}  

\newcommand{\code}[1]{\textsf{#1}} 
\newcommand{\bcode}[1]{\texttt{#1}}

\definecolor{lightgreen}{rgb}{.85,.95,.85}
\definecolor{lightblue}{rgb}{.85,.90,1}
\definecolor{lightred}{rgb}{.95,.85,.85}
\definecolor{lightgrey}{rgb}{.95,.95,.95}
\sethlcolor{lightred}

\definecolor{drkyellow}{rgb}{0.4,0.3,0}
\definecolor{drkred}{rgb}{0.5,0,0}
\definecolor{drkgreen}{rgb}{0,0.5,0}

\definecolor{dkgreen}{rgb}{0.1,0.40,0.1}

\definecolor{dkred}{rgb}{0.5,0,0}
\definecolor{medred}{rgb}{0.6,0,0}
\definecolor{gray}{rgb}{0.5,0.5,0.5}
\definecolor{mauve}{rgb}{0.58,0,0.82}

\newtheorem{algorithm}{Algorithm}

\newcommand{\aset}[1]{\{#1\}}
\newcommand{\dom}{\mathop\textit{dom}\nolimits}

\newcommand{\sfmt}[1]{\textsf{#1}}
\newcommand{\sch}{\textit{name}}
\newcommand{\loc}{\ell}
\newcommand{\sassign}[2]{#1 := #2}

\newcommand{\sderef}[1]{!#1}
\newcommand{\sfalse}{\sfmt{false}}
\newcommand{\sif}[3]{\sfmt{if}~#1~\sfmt{then}~#2~\sfmt{else}~#3}

\newcommand{\sinstall}[2]{\sfmt{install}~#1~#2}

\newcommand{\sref}[1]{\sfmt{ref}~#1}

\newcommand{\ssend}[2]{\sfmt{send}~#1~#2}
\newcommand{\strue}{\sfmt{true}}
\newcommand{\sunit}{\sfmt{unit}}
\newcommand{\sreduce}{\Downarrow}
\newcommand{\treduce}{\rightarrow}

\newcommand{\judge}{\vdash}
\newcommand{\xv}{p}

\newcommand{\traces}{\textit{traces}}

\newcommand{\atom}{A}
\newcommand{\toolname}{ClickRelease\xspace}

\newcommand{\tr}{t\xspace}

\newcommand{\tset}{\ensuremath{\mathcal{T}}\xspace}

\newcommand{\tnext}{\mathcal{X}}
\newcommand{\talways}{\mathcal{G}}

\newcommand{\tfuture}{\mathcal{F}}
\newcommand{\tuntil}{~\mathcal{U}~}
\newcommand{\tsince}{~\mathcal{S}~}
\newcommand{\tpast}{\mathcal{P}}

\newcommand{\tlast}[2]{\textit{last}(#1, #2)}
\newcommand{\limplies}{\rightarrow}

\newcommand{\evt}{\eta}
\newcommand{\mimplies}{\Rightarrow}

\newcommand{\tlevel}[3]{\textit{level}(#1, #2, #3)}
\newcommand{\tleveltr}[2]{\textit{level}(#1, #2)}

\begin{document}
\title{Checking Interaction-Based
  Declassification Policies for Android Using Symbolic
  Execution\thanks{This research was supported in part by NSF grants CNS-1064997 and CNS-1421373, 
AFOSR grants FA9550-12-1-0334 and FA9550-14-1-0334,
the partnership between UMIACS and the Laboratory for Telecommunication Sciences,
and the National Security Agency.}}

\author{Kristopher Micinski\inst{1} \and
  Jonathan Fetter-Degges\inst{1} \and
  Jinseong Jeon\inst{1} \and \\
  Jeffrey S. Foster\inst{1} \and
  Michael R. Clarkson\inst{2}}
\institute{University of Maryland, College Park \\
\email{\{micinski,jonfd,jsjeon,jfoster\}@cs.umd.edu}
\and 
Cornell University \\
\email{clarkson@cs.cornell.edu}
}

\maketitle

\begin{abstract} 
  Mobile apps can access a wide variety of secure information, such as
  contacts and location. However, current mobile platforms include
  only coarse access control mechanisms to protect such data.  In this
  paper, we introduce \emph{interaction-based declassification
    policies}, in which the user's interactions with the app constrain
  the release of sensitive information.  Our policies are defined
  extensionally, so as to be independent of the app's implementation,
  based on sequences of security-relevant events that occur in app
  runs. Policies use LTL formulae to precisely specify which secret
  inputs, read at which times, may be released. We formalize a
  semantic security condition, \emph{interaction-based
    noninterference}, to define our policies precisely.  Finally, we
  describe a prototype tool that uses symbolic execution of Dalvik bytecode to check
  interaction-based declassification policies for Android, and we show
  that it enforces policies correctly on a set of apps.
\end{abstract}

\section{Introduction}
\label{sec:introduction}

The Android platform includes a \emph{permission} system that aims to
prevent apps from abusing access to sensitive information, such as
contacts and location. Unfortunately, once an app is installed, it has
\emph{carte blanche} to use any of its permissions in arbitrary ways
at run time. For example, an app with location and Internet access
could continuously broadcast the device's location, even if such
behavior is not expected by the user.

To address this limitation, this paper presents a new framework for
Android app security based on \emph{information flow control}
\cite{Denning:1975} and user interactions.  The key idea behind our framework
is that users naturally express their intentions about information
release as they interact with an app.  For example,
clicking a button may permit an app to release a phone number over the
Internet. Or, as another example,
toggling a radio button from ``coarse'' to ``fine'' and back to
  ``coarse'' may temporarily permit an app to use fine-grained GPS
  location rather than a coarse-grained approximation.

To model these kinds of scenarios, we introduce
\emph{interaction-based declassification policies}, which
extensionally specify what information flows may occur after which
sequences of \emph{events}.  Events are GUI interactions (e.g.,
clicking a button), inputs (e.g., reading
the phone number), or outputs (e.g., sending over the Internet).
A policy is a set of \emph{declassification conditions}, written
$\phi \mathrel\rhd S$, where $\phi$ is a linear-time temporal logic
(LTL)~\cite{Pnueli:1977} formula over events, and $S$ is a sensitivity
level.  If $\phi$ holds at the time an input occurs, then that input
is declassified to level $S$. 
We formalize a semantic security condition,
\emph{interaction-based noninterference} (IBNI), over sets of event
\emph{traces} generated by an app.  Intuitively, IBNI holds of an app
and policy if observational
determinism~\cite{Zdancewic:03} holds after all inputs have been
declassified according to the policy. 
(Section~\ref{sec:overview}
describes policies further, and Section~\ref{sec:formalism} presents
our formal definitions.)

We introduce \toolname, a static analysis tool to check
whether an Android app and its declassification policy
satisfy IBNI.
\toolname generates event traces using SymDroid~\cite{Jeon:2012}, a
Dalvik bytecode symbolic executor.  \toolname
works by simulating user interactions with the app and recording the
resulting execution traces.
In practice, it is not feasible to
enumerate all program traces, so \toolname generates traces up to some
\emph{input depth} of $n$ GUI events.  
\toolname{} then synthesizes a
set of logical formulae that hold if and only if IBNI holds, and uses
Z3~\cite{deMoura:2008} to check their satisfiability.
(Section~\ref{sec:implementation} describes \toolname in detail.)

To validate \toolname, we used it to analyze four Android apps,
including both secure and insecure variants of those apps.
We ran each app variant under a range of input depths, and confirmed
that, as expected, \toolname{} scales exponentially.
However, we manually examined each app and its policy, and
found that an input depth of at most 5 is sufficient to
guarantee detection of a security policy violation (if any) for these
cases.  We ran \toolname{} at these minimum input depths and found
that it correctly passes and fails the secure and insecure app
variants, respectively. Moreover, at these depths, \toolname{} takes just a few
seconds to run. (Section~\ref{sec:experiments} describes our experiments.)

In summary, we believe that \toolname takes an important step forward
in providing powerful new security mechanisms for mobile devices. We
expect that our approach can also be used in other
GUI-based, security-sensitive systems.

\section{Example Apps and Policies}
\label{sec:overview}

We begin with two example apps that show interesting aspects of
interaction-based declassification
policies.

\paragraph*{Bump app.}

The boxed portion of Fig.~\ref{fig:app-bump} gives (simplified) source
code for an Android app that releases a device's unique ID and/or phone
number. This app is inspired by the Bump app, which let users tap
phones to share selected information with each other.  We have
interspersed an insecure variant of the app in the red code on
lines~\ref{line:bump1} and \ref{line:bump2}, which we will
discuss in Section~\ref{sec:traces}.

Each screen of an Android app is implemented using a class that
extends \code{Activity}. When an app is launched, Android invokes the
\code{onCreate} method for a designated main activity.
(This is part of the \emph{activity lifecycle}~\cite{Android:15},
which includes several methods called in a certain order. For this
simple app, and the other apps used in this paper, we only need a
single activity with this one lifecycle method.)
That method retrieves
(lines~\ref{line:bump-button}--\ref{line:bump-check}) the GUI IDs of a
button (marked ``send'') and
two checkboxes (marked ``ID'' and ``phone''). The \code{onCreate} method next
gets an instance of the \code{TelephonyManager}, uses it
to retrieve the device's unique ID and phone number information, and unchecks the two
checkboxes as a default. Then it creates a new callback
(line~\ref{line:bump-cb}) to be invoked when the ``send'' button is
clicked. When called, that callback releases the user's ID and/or
phone number, depending on the checkboxes.

\begin{figure}[t]
\centering
  \begin{lstlisting}[name=Ex]
public class BumpApp extends Activity {
 protected void onCreate(...) {
  Button sendBtn = (Button) findViewById(...); /**\label{line:bump-button}*/
  CheckBox idBox = (CheckBox) findViewById(...);
  CheckBox phBox = (CheckBox) findViewById(...);/**\label{line:bump-check}*/
  TelephonyManager manager = TelephonyManager.getTelephonyManager(); /**\label{line:bump-tele}*/
  final int id = manager.getDeviceId(); /**\label{line:bump-id}*/
  final int ph = manager.getPhoneNumber();
  idBox.setChecked(false); phBox.setChecked(false);
  sendBtn.setOnClickListener(
   new OnClickListener() { /**\label{line:bump-cb}*/
    public void onClick(View v) {
     if (idBox.isChecked())
      Internet.sendInt(id); /**\textcolor{red}{//Internet.sendInt(ph);}\label{line:bump1}*/
     if (phBox.isChecked())
      Internet.sendInt(ph); /**\textcolor{red}{//Internet.sendInt(id);}\label{line:bump2}*/
}})}}
  \end{lstlisting}
\begin{displaymath}
  \begin{array}{cc}
      \code{id}!\ast \wedge (\tfuture ( \code{sendBtn!unit} \land
      \tlast{\code{idBox}}{\code{true}})) \rhd Low, \\

      \code{ph}!\ast \wedge (\tfuture (
      \code{sendBtn!unit} \land
      \tlast{\code{phBox}}{\code{true}})) \rhd Low \\
    \end{array}
\end{displaymath}
\caption{``Bump'' app and policy.}
\label{fig:app-bump}
\end{figure}
 
This app is written to work with \toolname{}, a symbolic execution tool we built to check
whether apps satisfy interaction-based declassification policies. As we
discuss further in Section~\ref{sec:implementation}, \toolname{} uses an
executable model of Android that abstracts away some details that are
unimportant with respect to security. While a real app would release
information by sending it to a web server, here we instead call a
method \code{Internet.sendInt}. Additionally, while real apps
include an XML file specifying the screen layout of buttons,
checkboxes, and so on, \toolname{} creates those GUI elements
on demand at calls to \code{findViewById} (since their screen locations are
unimportant). Finally, we model the ID and phone number as
integers to keep the analysis simpler.

\toolname{} symbolically executes paths through subject apps, recording a
\emph{trace} of \emph{events} that correspond to certain method calls.
For example, one path through this app generates a trace
\begin{displaymath}
   \code{id!42}, \code{ph!43}, \code{idBox!true},
    \code{sendBtn!unit}, \code{netout!}\code{42}
\end{displaymath}
Each event has a \emph{name} and a \emph{value}. Here we have used
names \code{id} and \code{ph} for secret inputs, \code{idBox} and
\code{sendBtn} for GUI inputs, and \code{netout} for the network
send.  In particular, the trace above indicates 42 is read as the ID,
43 is read as the phone number, the ID checkbox is selected, the send
button is clicked (carrying no value, indicated by
\code{unit}), and then 42 is sent on the network. In \toolname{},
events are generated by calling certain methods that are specially
recognized. For example, \toolname{} implements the
\code{manager.getDeviceId} call as both returning a value and emitting
an event.

Notice here that in the trace, callbacks to methods such as
\code{idBox} and \code{sendBtn} correspond to user
interactions. The key idea behind our framework is that these actions
convey the user's intent as to which information should be
released. Moreover, traces also contain actions relevant to
information release---here the reads of the ID and phone number, and
the network send. Thus, putting both user interactions and
security-sensitive operations together in a single trace allows
our policies to enforce the user's intent.

The policy for
this example app is shown at the bottom of Fig.~\ref{fig:app-bump}.
Policies are comprised of a set of \emph{declassification
  conditions} of the form $\phi \rhd S$, where $\phi$ is an LTL
formula describing event traces and $S$ is a security level.  Such a
condition is read, ``At any input event, if $\phi$ holds at that
position of the event trace, then that input is declassified at level
$S$.''  For this app there are two declassification conditions. The
top condition declassifies (to \emph{Low}) an input that is a
read of the ID at any value (indicated by $\ast$), if
sometime in the future (indicated by the $\tfuture$ modality) the send
button is clicked and, when that button is clicked, the last value of
the ID checkbox was \code{true}. (Note that \emph{last} is not
primitive, but is a macro that can be expanded into regular LTL.)  The
second declassification condition does the analogous thing for the
phone number.

To check such a policy, \toolname{} symbolic executes the program,
generating per-path traces; determines the classification level of every input; and
checks that every pair of traces satisfies noninterference.
Note that using LTL provides a very general and
expressive way to describe the sequences of events that imply
declassification. For example, here we precisely capture that
only the last value of the checkbox matters for declassification. For
example, if a user selects the ID checkbox but then unselects it
and clicks send, the ID may not be released.

Although this example relies on a direct flow, \toolname{} can also
detect implicit flows. Section~\ref{sec:policies} defines an
appropriate version of noninterference, and the experiments in
Section~\ref{sec:experiments} include a subject program with an
implicit flow.


Notice this policy depends on the app reading the ID
and phone number when the app starts. If the app instead
waited until after the send button were clicked, it would violate this
policy. We could address this by replacing the $\tfuture$ modality by
$\tpast$ (past) in the policy, and we could form a disjunction of the
two policies if we wanted to allow either implementation. More generally, we
designed our framework to be sensitive to such choices to
support  reasoning about secret
values that change over time. We will see an
example next.

\paragraph*{Location resolution toggle app.}

\begin{figure}[t]
\centering
\begin{lstlisting}[name=Ex]
public class ToggleRes extends Activity { ...
 LocSharer mLocSharer = new LocSharer();
 RadioManager mRadio = new RadioManager();
 protected void onCreate(...) { ... }
 private class LocSharer implements LocationListener { ... 
  public LocSharer(RadioManager rm) {
   lm = (LocationManager) getSystemService(LOCATION_SERVICE);
   lm.requestLocationUpdates(mCurrentProvider, SHARE_INTERVAL, distance, this);
  }
  public void onLocationChanged(Location l) {
   if (mRadio.mFine) {
    Internet.sendInt(l.mLatitude);
    Internet.sendInt(l.mLongitude);
   } else {
    Internet.sendInt(l.mLatitude & 0xffffff00);
    Internet.sendInt(l.mLongitude & 0xffffff00);
 } } }
 private class RadioManager
  implements OnClickListener {
   public boolean mFine = false;
   public void onClick(View v) { mFine = !mFine; }
} }
\end{lstlisting}
  \begin{displaymath}
    \begin{array}{ll}
    \code{longitude}!\ast \wedge
    \tlast{\code{mRadio}}{\code{true}} \rhd
    \textit{Low}, \\
    \code{longitude}!\ast \wedge
    \tlast{\code{mRadio}}{\code{false}} \rhd
    \textit{MaskLower8}
    \end{array}
  \end{displaymath}
\caption{Location sharing app and policy.}
\label{fig:app-loc-toggle}
\end{figure}

Fig.~\ref{fig:app-loc-toggle} gives code for an app that
shares location information, either at full or truncated resolution
depending on a radio button setting. The app's \code{onCreate}
method displays a radio button (code not shown) and then creates and
registers a new instance of \code{RadioManager} to be called
each time the radio button is changed. That
class maintains field \code{mFine} as \code{true} when the radio button is
set to full resolution and \code{false} when set to truncated
resolution.

Separately, \code{onCreate} registers \code{LocSharer} to be called
periodically with the current location.  It requests location updates
by registering a callback with the \code{LocationManager}
system service.  When called, \code{LocSharer} releases the
location, either at full resolution or with the lower 8 bits
masked, depending on \code{mFine}.

The declassification policy for longitude appears below the code; the
policy for latitude is analogous.  This policy allows the precise
longitude to be released when
\code{mRadio} is set to fine, but only the lower eight bits to
be released if \code{mRadio} is set to coarse. Here \toolname{}
knows that at the \textit{MaskLower8} level, it should consider
outputs to be equivalent up to differences in the lower 8
bits. 

Finally, notice that this policy does not use the future
modality. This is deliberate, because location may be read multiple
times during the execution, at multiple values, and the security level
of those locations should depend on the state of the
radio button at that time. For example, consider a trace
\begin{displaymath}
   \code{mRadio!false}, \code{longitude!}v_1,
    \code{mRadio!true}, \code{longitude!}v_2
\end{displaymath}
The second declassification condition ($\code{longitude}!\ast \wedge
\tlast{\code{mRadio}}{\code{false}}$) will match the event with $v_1$, since
the last value of \code{mRadio} was \code{false}, and
thus $v_1$ may be declassified only to \textit{MaskLower8}. Whereas
the first declassification condition will match the event with $v_2$, hence it
may be declassified to \textit{Low}.

\section{Program Traces and Security Definition}
\label{sec:formalism}

Next, we formally define when a set of program traces satisfies an
interaction-based declassification policy.

\subsection{Program Traces}
\label{sec:traces}

\begin{figure}[t!]
  \small
  \centering
  \begin{displaymath}
    \begin{array}{p{1in}lrl}
      Primitives & \xv & ::= & n \mid \strue \mid \sfalse \mid \sunit \mid f(\xv_1, \ldots, \xv_i) \\
      Events & \evt & ::= & \sch ! p \\
      Traces & t & ::= & \evt ~ \textit{list} \\
      \\
      \multicolumn{4}{c}{\textrm{(a) Event and Trace Definitions.}} \\
      \\
      Policies & P & ::= & C_1, C_2, \ldots \\
      Conditions & C & ::= & \phi \rhd S\\
      Security Levels & S & ::= & \textit{High} \mid \textit{Low} \mid
      \textit{MaskLower8} \mid \ldots \\
      Atoms & \atom & ::= & \sch!s \mid s \oplus s \\
      Messages & s & ::= & x \mid p \mid \ast \\
      Formulae & \phi & ::= &
      \atom
      \mid \neg \phi
      \mid \phi \wedge \phi
      \mid \phi \vee \phi
      \mid \phi \limplies \phi
      \mid \exists x.\phi 
      \mid \forall x.\phi \\
      && \mid  & \tnext \phi
      \mid \phi \tuntil \phi
      \mid \talways \phi
      \mid \tfuture \phi
      \mid \phi \tsince \phi
      \mid \tpast \phi \\
      \\
      \multicolumn{4}{c}{\textrm{(b) Interaction-based Declassification Policy Language.}}
    \end{array}
  \end{displaymath}
  \caption{Formal definitions.}
  \label{fig:formalism}
\end{figure}

Fig.~\ref{fig:formalism}(a) gives the formal syntax of events and
traces.  \emph{Primitives}~$p$ are terms that can be carried by
events, e.g., values for GUI events, secret inputs, or network
sends.  In our formalism, primitives are integers, booleans, and terms
constructed from primitives using uninterpreted constructors $f$.  As
programs execute, they produce a \emph{trace}~$\tr$ of
\emph{events}~$\evt$, where each event $\sch!p$ pairs an event name
$\sch$ with a primitive $p$. We assume event names are partitioned
into those corresponding to inputs and those corresponding to
outputs. For all the examples in this paper, all names are inputs
except \code{netout}, which is an output.

Due to space limitations, we omit details of how traces are
generated.  These details, along with definition of our LTL 
formulas, can be found in appendices~\ref{sec:semantics} and
\ref{sec:ltl}, respectively.
Instead, we simply assume there exists some set
$\tset$ containing all possible traces a given program may
generate.
For example, consider the insecure variant bump app in
Fig.~\ref{fig:app-bump}, which replaces the black code with the red
code on lines lines \ref{line:bump1}
and \ref{line:bump2}.  This app sends the phone number when the
email box is checked and vice-versa. Thus, its set $\tset$
contains, among others, the following two traces:
\begin{displaymath}
  \begin{array}{cl}
    \code{id}!0, \code{ph}!0, \code{idBox}!\code{true},
    \code{sendBtn}!\code{unit}, \code{netout}!0 & (1) \\
    \code{id}!0, \code{ph}!1, \code{idBox}!\code{true},
    \code{sendBtn}!\code{unit}, \code{netout}!1 & (2) \\
  \end{array}
\end{displaymath}%
\lstset{language=Java}%
In the first trace, ID and phone number are read as 0, the
ID checkbox is selected, the button is clicked, and 0 is sent.
The second trace is similar, except the phone number and sent value
are 1. Below, we use these traces to show this program
violates its security policy.

\subsection{Interaction-based Declassification Policies}
\label{sec:policies}

We now define our policy language precisely.
Fig.~\ref{fig:formalism}(b) gives the formal syntax of
declassification policies.  A policy $P$ is a set of
\emph{declassification conditions} $C_i$ of the form $\phi_i\rhd S_i$,
where $\phi_i$ is an LTL formula describing when an input is
declassified, and $S_i$ is a \emph{security level} at which the value
in that event is declassified.

As is standard, security levels $S$ form a lattice.  For our
framework, we require that this lattice be finite.  We include
\textit{High} and \textit{Low} security levels, and we can generalize
to arbitrary lattices in a straightforward way. Here we include the
\textit{MaskLower8} level from Fig.~\ref{fig:app-loc-toggle} as an
example, where $\textit{Low} \sqsubseteq \textit{MaskLower8}
\sqsubseteq \textit{High}$.  Note that although we include
\textit{High} in the language, in practice there is no reason to
declassify something to level \textit{High}, since then it remains
secret.

The \emph{atomic predicates}~$A$ of LTL formulae match events,
e.g., atomic predicate $\sch!p$ matches exactly that event.
We include $\ast$ for matches to
arbitrary primitives. We allow event values to be
variables that are bound in an enclosing quantifier. The atomic
predicates also include atomic arithmetic
statements; here $\oplus$ ranges over standard operations such as $+$,
$<$, etc.  
The combination of these lets us describe complex
events. For example, we could write
$\exists x. \textit{spinner}!x \wedge x > 2$ to indicate the
\emph{spinner} was selected with a value greater than 2.

Atomic predicates are combined with the usual boolean connectives
($\neg$, $\wedge$, $\vee$, $\rightarrow$) and existential and
universal quantification.  Formulae include standard LTL
modalities $\tnext$ (next), $\tuntil$
(until), $\talways$ (always), $\tfuture$ (future), $\phi \tsince \psi$
(since), and $\tpast \phi$ (past). We include a wide range of
modalities, rather than a minimal set, to make policies easier to write.
Formulae also include
$\tlast{\sch}{p}$, which is syntactic sugar for $\lnot (\sch!\ast)
\tsince \sch!p$.
We assume a standard interpretation of LTL formulae over
traces \cite{Lichtenstein:85}.
We write $\tr, i \models \phi$ if trace $\tr$ is a model of $\phi$ at
position $i$ in the trace.




Next consider a trace $\tr \in \tset$ for an arbitrary program.
We write $\tlevel{\tr}{P}{i}$ for the security level that policy
$P$ assigns to the event $\tr[i]$:

\begin{displaymath}
  \tlevel{\tr}{P}{i} =
  \begin{cases}
    \bigsqcap_{\phi_j\rhd S_j \in P} \aset{ S_j \mid \tr, i \models
      \phi_j } & \tr[i] = \sch!p \\
    \textit{Low} & \tr[i] = \code{netout}!p \\
  \end{cases}
\end{displaymath}

In other words, for inputs, we take the greatest lower bound (the most
declassified) of the levels from all declassification conditions that
apply. We always consider network outputs to be
declassified. Notice that if no policy applies, the level is $H$ by
definition of greatest lower bound.

For example, consider trace (1) above with
respect to the policy in Fig.~\ref{fig:app-bump}.  At position 0, the
LTL formula holds because the ID box is eventually checked and then
the send button is clicked, so $\tlevel{(1)}{P}{0} =
\textit{Low}$. However,
$\tlevel{(1)}{P}{1} = \textit{High}$ because no
declassification condition applies for \code{ph}
(\code{phBox} is never checked). And $\tlevel{(1)}{P}{4} =
\textit{Low}$, because that position is a network send.

Next consider applying this definition to the GUI inputs. As written,
we have $\tlevel{(1)}{P}{2}$ = $\tlevel{(1)}{P}{3}$ =
\textit{High}. However, our app is designed to leak these inputs. 
For example, an adversary will learn the state of
\code{idBox} if they receive a message with an ID. Thus,
for all the subject apps in this paper, we also declassify all GUI inputs as
\textit{Low}. 
For the example in Fig.~\ref{fig:app-bump}, this means
adding the conditions
$\code{idBox!}\ast \rhd \textit{Low}$,
$\code{phBox!}\ast \rhd \textit{Low}$, and
$\code{sendBtn!}\ast \rhd \textit{Low}$. In general, 
the security policy designer should decide the security level of GUI inputs.

Next, we can apply \textit{level} pointwise across a trace and discard
any trace elements that are below a given level $S$. We define
\begin{displaymath}
\tleveltr{\tr}{P}^S[i] =
\begin{cases}
\tr[i] & \tlevel{\tr}{P}{i} \sqsubseteq S \\
\tau & \textrm{otherwise}
\end{cases}
\end{displaymath}
We write $\tleveltr{\tr}{P}^{S,in}$ for the same filtering, except
output events (i.e., network sends) are removed as well.
Considering the traces (1) and (2) again, we have

\begin{displaymath}
  \begin{array}{r@{ }c@{ }l}
    \tleveltr{(1)}{P}^\textit{Low} & = & \code{id}!0, \code{idBox}!\code{true},
    \code{sendBtn}!\code{unit}, \code{netout}!0 \\
    \tleveltr{(2)}{P}^\textit{Low} & = & \code{id}!0, \code{idBox}!\code{true},
    \code{sendBtn}!\code{unit}, \code{netout}!1 \\
    \tleveltr{(1)}{P}^\textit{Low,in} & = & \code{id}!0, \code{idBox}!\code{true},
    \code{sendBtn}!\code{unit} \\
    \tleveltr{(2)}{P}^\textit{Low,in} & = & \code{id}!0, \code{idBox}!\code{true},
    \code{sendBtn}!\code{unit} \\
  \end{array}
\end{displaymath}

Finally, we can define a program to satisfy noninterference if, for
every pair of traces such that the inputs at level $S$ are the same,
the outputs at level $S$ are also the same.
To account for generalized lattice levels such as \textit{MaskLower8},
we also need to treat events that are equivalent at a certain level as
the same. For example, at \textit{MaskLower8}, outputs
\bcode{0xffffffff} and \bcode{0xffffff00} are the same, since they do
not differ in the upper 24 bits. Thus, we assume for each security
level $S$ there is a appropriate equivalence relation $=_S$, e.g., for
\textit{MaskLower8}, it compares elements ignoring their lower 8
bits. Note that $x =_\textit{Low} y$ is simply $x = y$ and
$x =_\textit{High} y$ is always true.

\begin{definition}[Interaction-based Noninterference (IBNI)]
  \label{defn:noninterference}
  A program satisfies security policy $P$, if for all $S$ and for
  all $t_1, t_2 \in
      \tset$ (the set of traces of the program) the following holds:
\begin{displaymath}
    \tleveltr{\tr_1}{P}^{S,in} =_S \tleveltr{\tr_2}{P}^{S,in}
    \implies
    \tleveltr{\tr_1}{P}^S =_S \tleveltr{\tr_2}{P}^S \\
\end{displaymath}
\end{definition}

Looking at traces for the insecure app, we see
they violate non-interference, because
$\tleveltr{(1)}{P}^\textit{Low,in} =
\tleveltr{(2)}{P}^\textit{Low,in}$, but
$\tleveltr{(1)}{P}^\textit{Low} \neq \tleveltr{(2)}{P}^\textit{Low}$
(they differ in the output).  We note that our definition of
noninterference makes it a 2-hypersafety property \cite{Clarkson:10,Clarkson:2014}.

\section{Implementation}
\label{sec:implementation}

We built a prototype tool, \toolname{}, to check whether Android apps obey the
interaction-based declassification policies described in
Section~\ref{sec:formalism}. \toolname{} is based on
SymDroid~\cite{Jeon:2012}, a symbolic executor for Dalvik bytecode,
which is the bytecode format to which Android apps are compiled.
As is standard, SymDroid computes with \emph{symbolic
  expressions} that may contain \emph{symbolic variables}
representing sets of values. At conditional branches that depend on
symbolic variables, SymDroid invokes Z3~\cite{deMoura:2008} to
determine whether one or both branches are feasible. As it follows
branches, SymDroid extends the current \emph{path condition}, which tracks
branches taken so far, and forks execution when multiple paths are
possible. Cadar and Sen~\cite{Cadar:13} describe
symbolic execution in more detail.

SymDroid uses the features of symbolic execution to implement
nondeterministic event inputs (such as button clicks or spinner
selections), up to a certain bound. Since we have symbolic variables
available, we also use them to represent arbitrary secret inputs, as
discussed below in Sec.~\ref{sec:symbolic-traces}. There are several issues that arise in applying SymDroid
to checking our policies, as we discuss next.

\subsection{Driving App Execution}
\label{sec:driver}

Android apps use the Android framework's API, which includes
classes for responding to events via callbacks. We could try to
account for these callbacks by symbolically execution Android framework code
directly, but past experience suggests this is intractable: the
framework is large, complicated, and includes native code.
Instead, we created an \emph{executable model}, written in Java, that
mimics key portions of Android needed by our subject apps. Our Android
model includes facilities for generating clicks and
other GUI events (such as the \code{View}, \code{Button}, and
\code{CheckBox} classes, among others). It also includes code for
\code{LocationManager},
\code{TelephonyManager}, and other basic Android classes.

In addition to code modeling Android, the model also
includes simplified versions of Java library classes such as
\code{StringBuffer} and \code{StringBuilder}.  Our versions of
these APIs implement unoptimized versions of methods in
Java and escape to internal SymDroid functions to handle operations that
would be unduly complex to symbolically execute. For instance, SymDroid
represents Java \code{String} objects with OCaml strings instead of
Java arrays of characters. It thus models methods such as \code{String.concat}
with internal calls to OCaml string manipulation functions. Likewise,
reflective methods such as \code{Class.getName} are handled internally.

For each app, we created a driver that uses our Android model to simulate user
input to the GUI. The driver is specific to the app since it depends on the
app's GUI.  The driver begins by calling the app's \code{onCreate}
method. 
Next it invokes special
methods in the Android model to inject GUI events. There is one such method for
each type of GUI element, e.g., buttons, checkboxes, etc. 
For example,
\code{Trace.addClick(id)} generates a click event for the given
\code{id} and then calls the appropriate event handler.
The trace entry contains the event name for that kind of element,
and a value if necessary. 
Event handlers are those
that the app registered through standard Android framework mechanisms,
e.g., in \code{onCreate}.

Let $m$ be the number of possible GUI events.  To simulate one
arbitrary GUI event, the driver uses a block that branches $m$ ways on
a fresh symbolic variable, with a different GUI action in each branch.
Typical Android apps never exit unless the framework kills them, and
thus we explore sequences of events only up to a user-specified
\emph{input depth}~$n$. Thus, in total, the driver will execute
at least $m^n$ paths.

\subsection{Symbolic Variables in Traces}
\label{sec:symbolic-traces}

In addition to GUI inputs, apps also use secret inputs. We could use
SymDroid to generate concrete secret inputs, but instead we opt to use
a fresh symbolic variable for each secret input. For example, the call
to \code{manager.getDeviceId} in Fig.~\ref{fig:app-bump} returns a
symbolic variable, and the same for the call to
\code{manager.getPhoneNumber}. This choice makes checking policies
using symbolic execution a bit more powerful, since, e.g., a symbolic
integer variable represents an arbitrary 32-bit integer. Note that
whenever \toolname generates a symbolic variable for a secret input, it
also generates a trace event corresponding to the input.

Recall that secret inputs may appear in traces, and thus traces may
now contain symbolic variables. For example, using $\alpha_i$'s as
symbolic variables for the secret ID and phone number inputs, the
traces (1) and (2) become

\begin{displaymath}
  \begin{array}{cl}
    \code{id}!\alpha_1, \code{ph}!\alpha_2, \code{idBox}!\code{true},
    \code{sendBtn}!\code{unit}, \code{netout}!\alpha_2 & (1') \\
    \code{id}!\alpha_1, \code{ph}!\alpha_2, \code{idBox}!\code{true},
    \code{sendBtn}!\code{unit}, \code{netout}!\alpha_2 & (2') \\
  \end{array}
\end{displaymath}

We must take care when symbolic variables are in traces.
Recall \textit{level} checks $t,i \models \phi$ and
then assigns a security level to position $i$. If $\phi$
depends on symbolic variables in $t$, we may not be able to
decide this. For example, if the third element in $(1')$ were
$\code{idBox}!\alpha_3$, then we would need to reason with
conditional security levels such as
$\tlevel{\tr}{P}{0} =\textsf{\textbf{if }} \alpha_3 \textsf{\textbf{ then }} \textit{Low}
\textsf{\textbf{ else }} \textit{High}$. We
avoid the need for such reasoning by only using symbolic variables for
secret inputs, and by ensuring the level assigned by a policy does not
depend on the value of a secret input. We leave supporting more complex
reasoning to future work.

\subsection{Checking Policies with Z3}

Each path explored by SymDroid yields a pair $(t, \Phi)$, where $t$ is
the trace and $\Phi$ is the path condition. \toolname{} uses Z3 to check whether a given set
of such trace--path condition pairs satisfies a policy $P$. Recall that
Definition~\ref{defn:noninterference} assumes for each $S$ there is an
$=_S$ relation on traces. We use the same relation below, encoding it
as an SMT formula. For our example lattice, $=_\textit{High}$ produces
\code{true}, $=_\textit{Low}$ produces a conjunction of equality tests
among corresponding trace elements, and $=_\textit{MaskLower8}$
produces the conjunction of equality tests of the bitwise-and of every
element with \bcode{0xffffff00}.

Given a trace
$t$, let $t'$ be $t$ with its symbolic variables primed, so that the
symbolic variables of $t$ and $t'$ are disjoint. Given a path
condition $\Phi$, define $\Phi'$ similarly. Now we can give the
algorithm for checking a security policy.

\begin{algorithm}
  To check a set $\tset$ of trace--path condition pairs, do the
  following. Let $P$ be the app's security policy. Apply \emph{level}
  across each trace to obtain the level of each event.  For each
  $(t_1, \Phi_1)$ and $(t_2, \Phi_2)$ in $\tset\times\tset$, and for
  each $S$, ask Z3 whether the following formula (the negation of
  Definition~\ref{defn:noninterference}) is unsatisfiable:
  \begin{displaymath}
      \tleveltr{\tr_1}{P}^{S,in} =_S \tleveltr{\tr_2'}{P}^{S,in} \land
      \tleveltr{\tr_1}{P}^S \neq_S \tleveltr{\tr_2'}{P}^S \land
      \Phi_1 \land \Phi_2' \\
  \end{displaymath}
  If no such formula is unsatisfiable, then the program satisfies noninterference.
\end{algorithm}
We include $\Phi_1$ and $\Phi'_2$ to
constrain the symbolic variables in the trace. More precisely,
$\tr_1$ represents a \emph{set} of concrete traces in which its symbolic
variables are instantiated in all ways that satisfy $\Phi_1$,
and analogously for $\tr'_2$.

If the above algorithm finds an unsatisfiable formula, then Z3 returns a counterexample, which
SymDroid uses in turn to generate a pair of concrete traces
as a counterexample.
For example, consider traces (1') and (2') above, and prime
symbolic variables in (2'). Those traces have the trivial path
condition \sfmt{true}, since neither branches on a symbolic
input. Thus, the formula passed to Z3 will be:
\begin{displaymath}
    \alpha_1 = \alpha'_1 \land \code{true} = \code{true} \land \sunit = \sunit
    \land
    \big(\alpha_1 \neq \alpha'_1 \vee \code{true} \neq \code{true} \vee
    \sunit \neq \sunit \vee \alpha_2 \neq \alpha'_2 \big)
\end{displaymath}
Thus we can see a satisfying
assignment with $\alpha_1 = \alpha'_1$ and $\alpha_2 \neq \alpha'_2$,
hence noninterference is violated.

\subsection{Minimizing Calls to Z3}
\label{sec:z3-tree}

A naive implementation of the noninterference check generates $n^2$
equations, where $n$ is the number of traces produced by \toolname{}
to be checked by Z3. However, we observed that many of these equations
correspond to pairs of traces with different sequences of GUI
events. Since GUI events are low inputs in all our policies, these
pairs trivially satisfy noninterference (the left-hand side of the
implication in Definition~\ref{defn:noninterference} is false).
 Thus, we need not send those
equations to Z3 for an (expensive) noninterference check.

We exploit this observation by organizing SymDroid's output traces
into a tree, where each node represents an event, with
the initial state at the root. Traces with common prefixes share the
same ancestor traces in the tree. We systematically traverse this tree
using a cursor $t_1$, starting from the root. When $t_1$ reaches a new
input event, we then traverse the tree using another cursor $t_2$,
also starting from the root. As $t_2$ visits the tree, we do not
invoke Z3 on any traces with fewer input events than $t_1$ (since they
are not low-equivalent to $t_1$). We also skip any subtrees where 
input events differ.

\section{Experiments}
\label{sec:experiments}

To evaluate \toolname{}, we ran it on four apps, including the two
described in Section~\ref{sec:overview}. We also ran \toolname{} on
several insecure variants of each app, to ensure it can detect the
policy violations. The apps and their variants
are:

\begin{itemize}[leftmargin=*]
\item \textit{Bump.} The bump app and its policy appear in
  Fig.~\ref{fig:app-bump}. The first insecure variant counts clicks to
  the send button sends the value of the
  ID after three clicks, regardless of the state of the ID
  checkbox. The second (indicated in the comments in the program text)
  swaps the released information---if the ID
  box is checked, it releases the phone number, and vice-versa.

\item \textit{Location toggle.} The location toggle app and
  its policy appear in Fig.~\ref{fig:app-loc-toggle}. The first insecure
  variant always shares fine-grained location information, regardless
  of the radio button setting. The second checks if coarse-grain
  information is selected. If so, it stores the fine-grained location
  (but does not send it yet).  If later the fine-grained radio button
  is selected, it sends the stored location. Recall this is forbidden
  by the app's security policy, which allows the release only of locations
  received while the fine-grained option is set.

\item \textit{Contact picker.} We developed a contact picker app
  that asks the user to select a contact from a spinner and then
  click a send button to release the selected contact information over the
  network. The security policy for this app requires that no contact
  information leaks unless it is the last contact selected before the
  button click. (For example, if the user selects contact 1,
  selects contact 2, and then clicks the button, only contact 2 may be
  released.) Note that since an arbitrarily sized list of contacts
  would be difficult for symbolic execution (since then there would be
  an unbounded number of ways to select a contact), we limit the app
  to a fixed set of three contacts.
  The first insecure variant of this app scans the set of contacts for a
  specific one. If found, it sends a message revealing that contact
  exists before sending the actual selected contact. The second insecure
  variant sends a different contact than was selected.

\item \textit{WhereRU.} Lastly, we developed an app
  that takes push requests for the user's location and shares it depending
  on user-controlled settings.
  The app contains a radio group with three buttons, ``Share Always,'' ``Share
  Never,'' and ``Share On Click.'' There is also a ``Share Now'' button that
  is enabled when the ``Share On Click'' radio button is selected.  When a
  push request arrives, the security policy allows sharing if (1) the ``Always''
  button is selected, or (2) the ``On Click'' button is selected and the user
  presses ``Share Now.'' Note that, in the second case, the location may
  change between the time the request arrives and the time the user authorizes
  sharing; the location to be shared is the one in effect when the user
  authorized sharing, i.e., the one from the most recent location update
  before the button click. Also, rather than include the full
  Android push request API in our model, we simulated it using a basic callback.
  This app has two insecure variants. In the first one, when the user
  presses the ``Share Now'' button, the app begins continuously
  sharing (instead of simply sharing the single location captured on
  the button press).  In the second, the app
  shares the location immediately in response to all requests.

\end{itemize}

\paragraph*{Scalability.}
We ran our experiments on a 4-core i7 CPU @3.5GHz with 16GB
RAM running Ubuntu 14. For each experiment we report the median of 10
runs.

\begin{figure}[t!]
\centering
\begin{tabular}{cc}
\includegraphics[width=.45\textwidth]{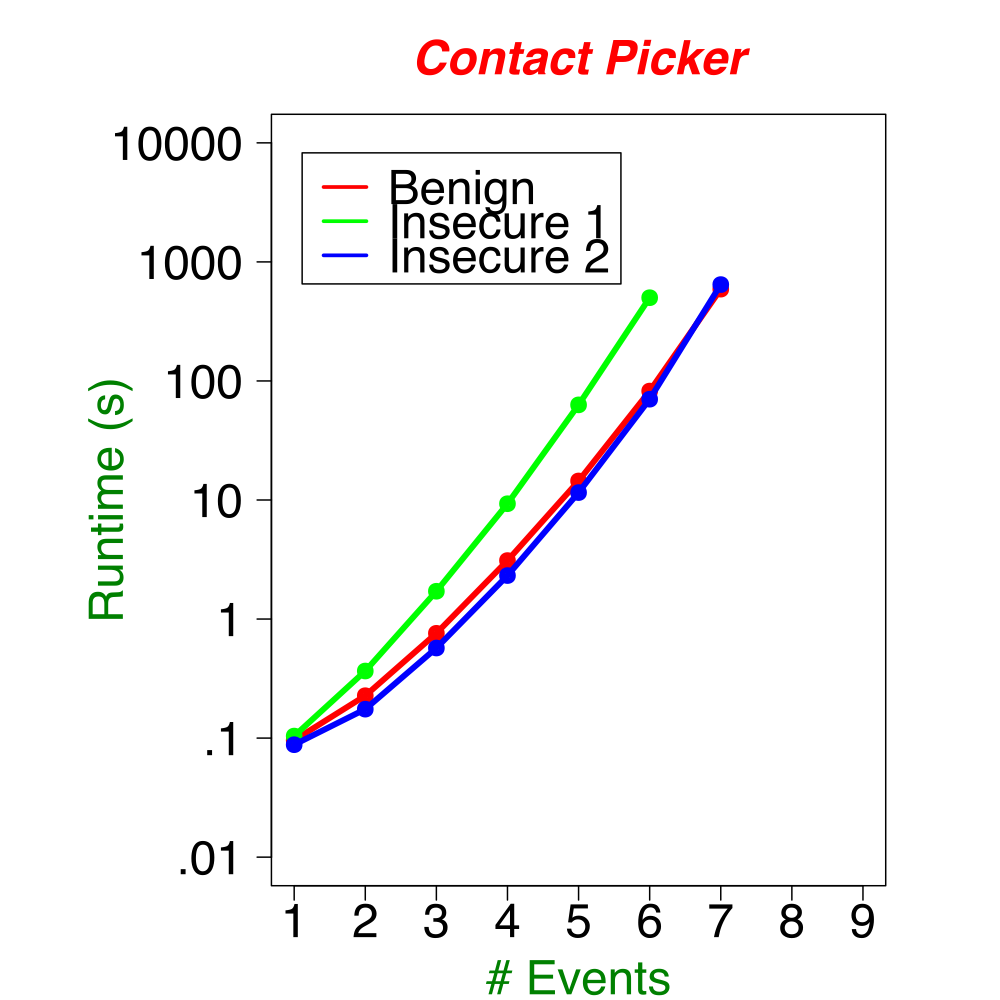} &
\includegraphics[width=.45\textwidth]{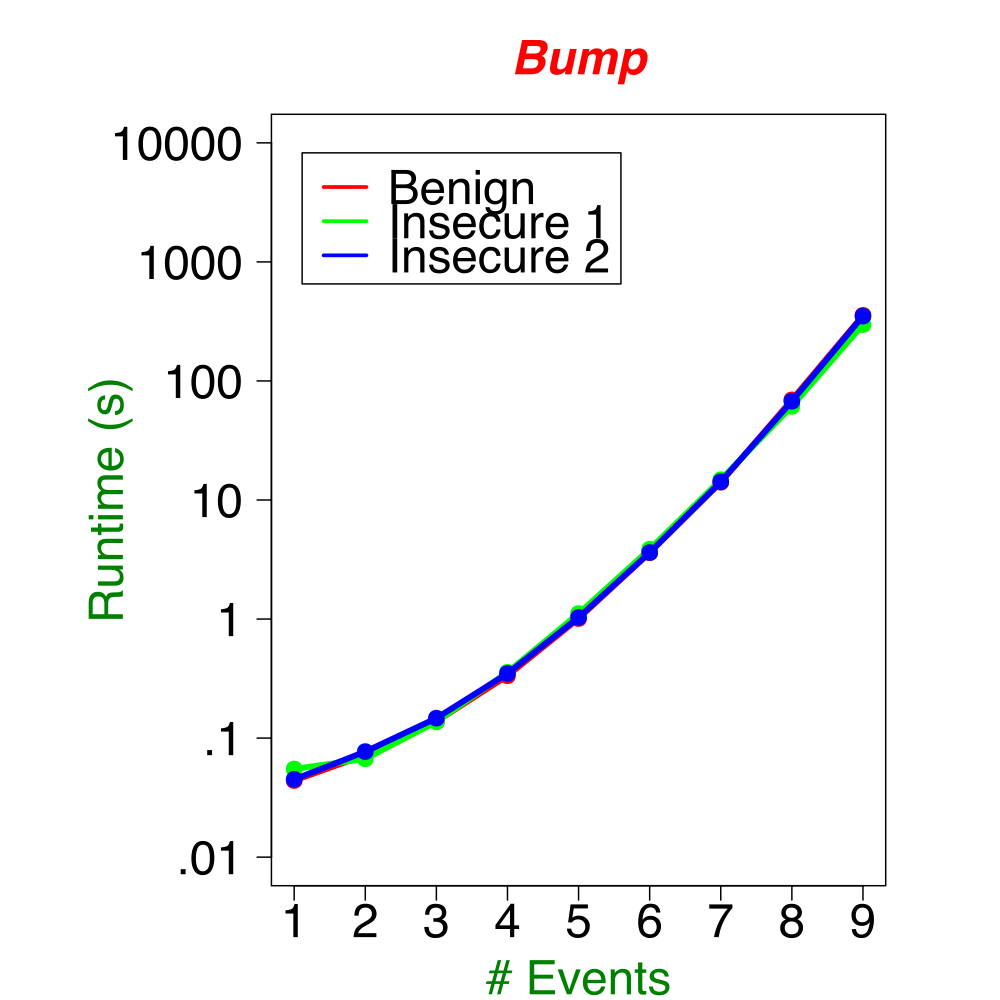} \\

\includegraphics[width=.45\textwidth]{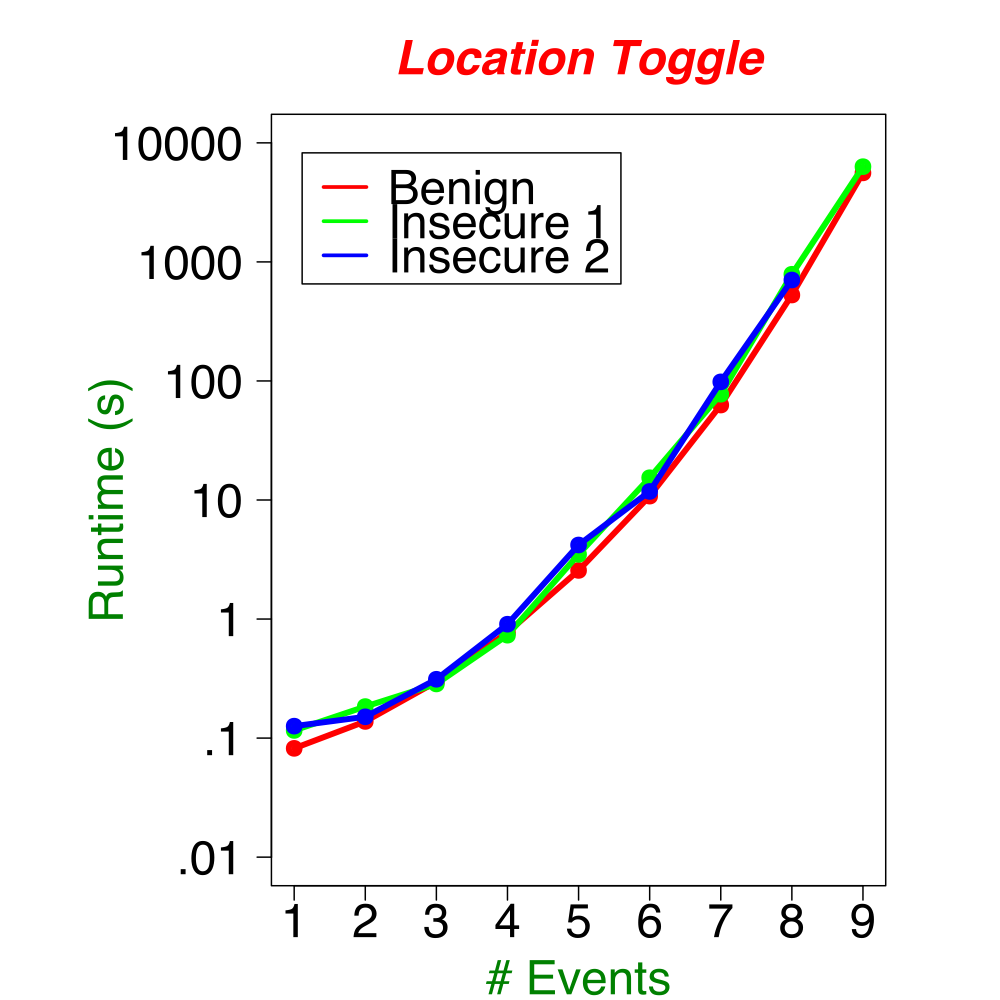} &
\includegraphics[width=.45\textwidth]{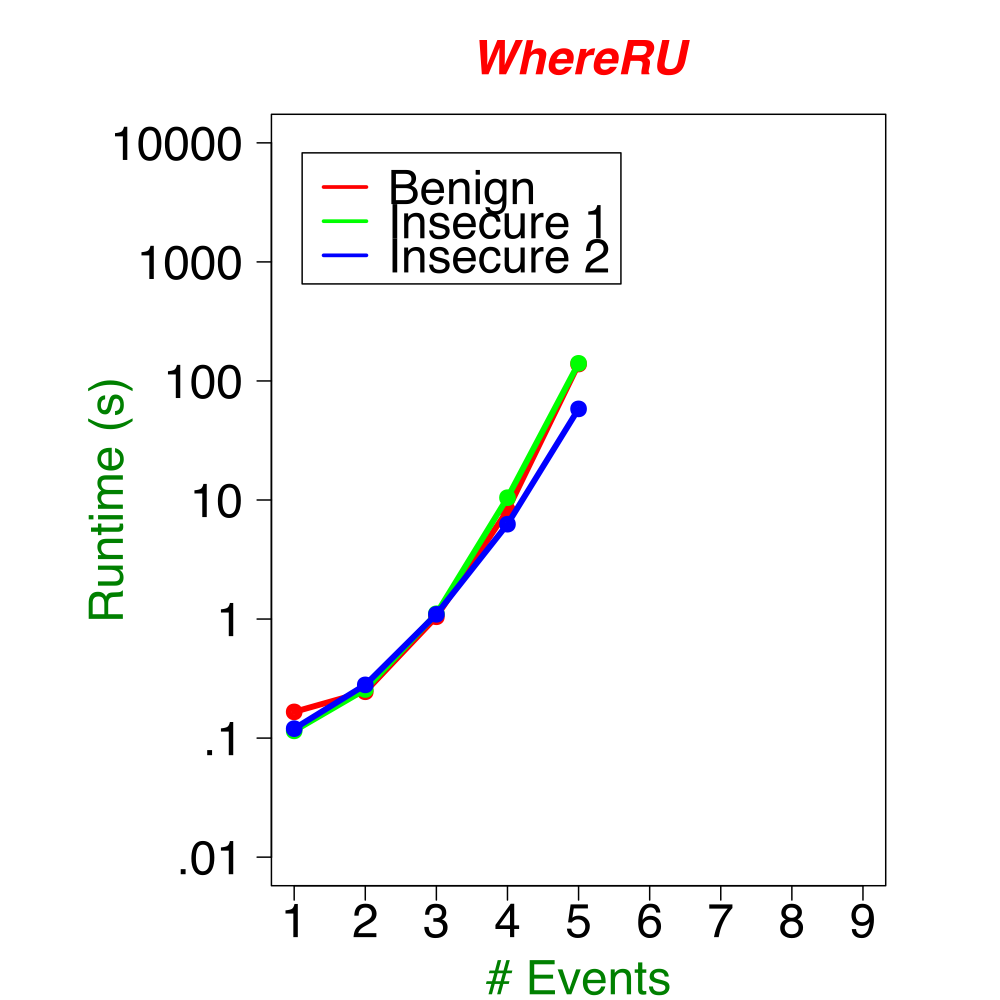}
\end{tabular}
\caption{Runtime vs. number of events.}
\label{fig:performance-graphs}
\end{figure}

In our first set of experiments, we measured how
\toolname{}'s performance varies with input
depth. Figure~\ref{fig:performance-graphs} shows running time (log scale)
versus input depth for all programs and variants.
For each app, we ran to the highest input depth that completed in
one hour.

For each app, we see that running time
grows exponentially, as expected. The maximum input depth
before timeout (i.e., where each curve ends) ranges from five to
nine. The differences have to do with the number of possible events at
each input point. For example, WhereRU has seven possible input events, so
it has the largest possible ``fan out'' and times out with an input
depth of five. In contrast, Bump and Location Toggle have just three
input events and time out with an input depth of nine. Notice also the first
insecure variant of Contact Picker times out after fewer events than
the other variants. Investigating further, this occurs due to
that app's implicit flow (recall the app branches on the value of a
secret input). Implicit flows cause symbolic execution to take
additional branches depending on the (symbolic) secret value.

\paragraph*{Minimum Input Depth.}

\begin{figure*}[t]
\centering
\begin{tabular}{ | l | r | r | r | r | r | }
  \hline
  &Input&\multicolumn{3}{c|}{Time (ms)} \\ \cline{3-5}
App & Depth & Exploration & Analysis & Total \\ \hline
  \hline
  Bump & 3 & 114 &  15 & 142 \\
  Bump (insecure 1) & 5 & 2,100 & 1,577 & 3,690 \\
  Bump (insecure 2) & 4 & 266 & 70 & 344 \\\hline
  Location toggle & 2 &  113 & 12 & 128 \\
  Location toggle (insecure 1) & 2 &  143 & 12 & 163 \\
  Location toggle (insecure 2) & 3 & 117 & 12 & 143 \\\hline
  Contact picker & 2 &  79 & 2 & 94 \\
  Contact picker (insecure 1) & 2 &  325 & 27 & 361 \\
  Contact picker (insecure 2) & 2 &  149 & 9 & 170 \\\hline
  WhereRU & 3 & 849 & 183 & 1,045 \\
  WhereRU (insecure 1) & 3 & 860 & 234 & 1,108 \\
  WhereRU (insecure 2) & 2 & 257 & 10 & 280 \\
  \hline
\end{tabular}
\caption{Results at minimum input depth.}
\label{fig:results}
\end{figure*}

Next, for each variant, we manually
calculated a \emph{minimum} input depth guaranteed
to find a policy violation.
To do so, first we determined possible app GUI states.
For example, in Bump (Fig.~\ref{fig:app-bump}), there is a state with \code{idBox} and
\code{phBox} both checked, a state with just \code{idBox} checked,
etc. Then we examined the policy and recognized that certain input
sequences lead to equivalent states modulo the policy.  For example,
input sequences that click \code{idBox} an even number of times and
then click send are all equivalent. Full analysis reveals that an
input depth of three (which allows the checkboxes to be set any
possible way followed by a button click) is
sufficient to reach all possible states for this policy. We performed
similar analysis on other apps and variants.

Fig.~\ref{fig:results} summarizes the results of running with
the minimum input depth for each variant, with the depths listed in
the second column.
We confirmed that, when run with this input depth, \toolname{}
correctly reports the benign app variants as secure and the other app
variants as insecure.
The remaining columns of Fig.~\ref{fig:results} report \toolname{}'s
running time (in milliseconds), broken down by the
exploration phase (where SymDroid generates the set of symbolic
traces) and the analysis phase (where SymDroid forms equations about
this set and checks them using Z3).  Looking at the breakdown between
exploration and analysis, we see that the former dominates the running
time, i.e., most of the time is spent simply exploring program
executions.  We see the total running time is typically around a
second or less, while for the first insecure variant of Bump it is
closer to 4 seconds, since it uses the highest input depth.

Our results show that while \toolname{} indeed scales exponentially,
to actually find security policy violations we need only run it with a
low input depth, which takes only a small amount of time.

\section{Limitations and Future Work}

There are several limitations of \toolname{} we plan to address
in future work.

Thus far we have applied \toolname{} to a set of small apps that we
developed. There are two main engineering challenges in applying
\toolname{} to other apps. First, our
model of Android (Section~\ref{sec:driver}) only includes part of the
framework. To run on other apps, it will need to be expanded
with more Android APIs. 
Second, we speculate
that larger apps may require longer input depths to go from app launch
to interfering outputs. In these cases, we may be able to start
symbolic execution ``in the middle'' of an app (e.g., as in the
work of Ma et al. \cite{Ma:2011}) to skip uninteresting
prefixes of input events.

\toolname{} also has several limitations related to its policy
language. First, \toolname{} policies are fairly low level. Complex
policies---e.g., in which clicking a certain button releases
multiple pieces of information---can be expressed, but are not very
concise. We expect as we gain more experience writing \toolname{}
policies, we will discover useful idioms that should be incorporated
into the policy language.  Similarly, situations where several
methods in sequence operate on and send information should be
supported.  Second, currently \toolname{} assumes there
is a single adversary who watches \code{netout}. It should be
straightforward to generalize to multiple output channels and multiple
observers, e.g., to model inter-app communication.  Third, we do not
consider deception by apps, e.g., we assume the policy writer knows
whether the \code{sendBtn} is labeled appropriately as ``send'' rather
than as ``exit.'' We leave looking for such deceptive practices to
future work.

Finally, since \toolname{} explores a limited number of program paths
it is not sound, i.e., it cannot guarantee the absence of policy
violations in general. However, in our experiments we were able to
manually analyze apps to show that exploration up to a certain input
depth was sufficient for particular apps, and we plan to investigate
generalizing this technique in future work.

\section{Related Work}
\label{sec:related-work}

\toolname is the first system to enforce extensional declassification policies 
in Android apps.  It builds on a rich history of research in usable security,
information flow, and declassification.

One of the key ideas in \toolname is that GUI interactions indicate the 
security desires of users.
Roesner et al.~\cite{Roesner:12} similarly propose \emph{access control gadgets}
(ACGs), which are GUI elements that, when users interact with them,
grant permissions. 
Thus, ACGs and \toolname both aim to better align 
security with usability~\cite{Yee:04}.
\toolname addresses secure information flow, especially propagation of
information after its release, whereas ACGs address only access control.

\paragraph*{Android-based systems.}

TaintDroid~\cite{Enck:10} is a run-time information-flow tracking system
for Android.  It monitors the usage of sensitive information and detects
when that information is sent over insecure channels.
Unlike \toolname{},
TaintDroid does not detect implicit flows.

AppIntent~\cite{Yang:2013} uses symbolic execution to derive the \emph{context},
meaning inputs and GUI interactions, that causes sensitive information to be
released in an Android app. A human analyst examines that context and makes an
expert judgment as to whether the release is a security violation.  
\toolname instead uses human-written LTL formulae to specify whether 
declassifications are permitted. It is unclear from~\cite{Yang:2013} whether AppIntent detects
implicit flows.

Pegasus~\cite{Chen:13} combines static analysis, model checking,
and run-time monitoring to check whether an app uses API
calls and privileges consistently with users' expectations.
Those expectations are expressed using LTL formulae, similarly to \toolname.
Pegasus synthesizes
a kind of automaton called a \emph{permission event graph} from the
app's bytecode then checks whether that automaton is a model for the formulae.
Unlike \toolname, Pegasus does not address information flow.

Jia et al.~\cite{Jia:13} present a system, inspired by Flume~\cite{Krohn:2007},
for run-time enforcement of information flow policies at the granularity of 
Android components and apps.  Their system allows
components and apps to perform trust declassification according to 
capabilities granted to them in security labels.  In contrast,
\toolname reasons about declassification in terms of user
interactions.

\paragraph*{Security type systems}

Security type systems \cite{Volpano:1996} statically disallow programs
that would leak information. O'Neill et al.~\cite{O'Neill:06} and
Clark and Hunt~\cite{Clark:09} define interactive variants of
noninterference and present security type systems that are sound with
respect to these definitions.

Integrating declassification with security type systems has been the
focus of much research.  Chong and Myers~\cite{Chong:04} introduce
\emph{declassification policies} that conditionally downgrade security labels.
Their policies use classical propositional logic for the conditions.  
\toolname can be seen as providing a more expressive language for
conditions by using LTL to express formulae over events.  
SIF (Servlet Information Flow)~\cite{Chong:07} is a framework for
building Java servlets with information-flow control.  Information
managed by the servlet is annotated in the source code with security
labels, and the compiler ensures that information propagates in ways
that are consistent with those labels.  The SIF compiler is based on
Jif \cite{Myers:1999}, an information-flow variant of Java.

All of these systems require adding type annotations to terms in the
program code, e.g., method parameters, etc. In contrast, \toolname{}
policies are described in terms of app inputs and
outputs.

\paragraph*{Event Based Models and Declassification}

Vaughan and Chong~\cite{Vaughan:2011} define expressive declassification policies that
allow functions of secret information to be released after events occur, and
extend the Jif compiler to infer events.  \toolname instead 
ties events to user interactions.

Rafnsson et al.~\cite{Rafnsson:12} investigate models, definitions, and enforcement
techniques for secure information flow in interactive programs in a
purely theoretical setting.
Sabelfeld and Sands~\cite{Sabelfeld:2009} survey approaches to
secure declassification in a language-based setting.  \toolname
can be seen as addressing their ``what'' and ``when'' axes of
declassification goals:  users of Android apps interact with the GUI
to control when information may be released, and the GUI is responsible
for conveying to the user what information will be released.

\section{Conclusion}
\label{sec:conclusion}

We introduced interaction-based declassification policies, which describe
\emph{what} and \emph{when} information can flow. Policies are defined
using LTL formulae describing event traces, where events include GUI
actions, secret inputs, and network sends. We formalized our policies
using a trace-based model of apps based on security relevant events.
Finally, we described \toolname{}, which uses symbolic
execution to check interaction-based declassification policies on Android, and
showed that \toolname{} correctly enforces policies on four apps,
with one secure and two insecure variants each.

\bibliographystyle{splncs03}
\bibliography{paper}

\appendix 

\section{Operational Semantics / Trace Generation}
\label{sec:semantics}

To illustrate how traces are generated, we introduce an operational
semantics for a model of Android programs.  Our semantics uses an
abstract machine which transitions between states and reads messages
on a message queue.  A \emph{state} is a tuple $(M, \sigma, H)$ that
includes a \emph{message queue}~$M$, which is a list of pairs of a channel and a primitive; a
\emph{heap}~$\sigma$, which maps locations to values; and a \emph{handler
  map}~$H$, which maps channel names to functions installed to
handle events on those channels. Each channel may have at most one handler.
We write $\Sigma.X$ (where $X$ could be $M$,
$\sigma$, or $H$), to mean the $X$ component of
$\Sigma$. Similarly, we write $\Sigma[X\mapsto X']$ to mean $\Sigma$
with the $X$ component replaced by $X'$.

As programs execute, they produce a \emph{trace}~$\tr$ of 
\emph{events}~$\evt$. Events are writes,
written $\sch ! p$, of primitive $p$ from channel $\sch$. We also include an empty event
$\tau$, which is the identity of trace concatenation.

\begin{figure*}[!t]
  \small
  \begin{displaymath}
    \begin{array}{c}
      \multicolumn{1}{l}{
        \framebox{$e, \Sigma_1 \sreduce v, \Sigma_2$}
      }
      \\ \\

      \infer[RVal]
      { }
      { v, \Sigma \sreduce v, \Sigma }

      \qquad

      \infer[RApp]
      {
        e_1, \Sigma_1 \sreduce (\lambda x.e_3), \Sigma_2 \\
        e_2, \Sigma_2 \sreduce v_1, \Sigma_3 \\\\
        e_3\aset{x\mapsto v_1}, \Sigma_3 \sreduce v_2, \Sigma_4
      }
      { e_1\;e_2, \Sigma_1 \sreduce v_2, \Sigma_4}

      \qquad

      \infer[RRef]
      {e, \Sigma_1 \sreduce v, \Sigma_2 \\
        \loc \not\in \dom(\Sigma_2.\sigma) \\\\
        \sigma' = (\Sigma_2.\sigma)[\loc\mapsto v] 
      }
      {\sref e, \Sigma \sreduce \loc, \Sigma_2[\sigma \mapsto \sigma']}

      \\ \\

      \infer[RAssign]
      {e_1, \Sigma_1 \sreduce \loc, \Sigma_2 \\
        \loc \in \dom(\Sigma_2.\sigma) \\\\
        e_2, \Sigma_2 \sreduce v, \Sigma_3 \\
        \sigma' = (\Sigma_3.\sigma)[\loc \mapsto v]
      }
      {\sassign {e_1} {e_2}, \Sigma_1 \sreduce
        v, \Sigma_3[\sigma \mapsto \sigma']}

      \qquad

      \infer[RDeref]
      {e, \Sigma_1 \sreduce \loc, \Sigma_2 \\
       (\Sigma_2.\sigma)(\loc) = v }
      {\sderef e, \Sigma_1 \sreduce v, \Sigma_2 }

      \\ \\

      \infer[RIfTrue]
      {e_1, \Sigma_1 \sreduce \code{true}, \Sigma_2 \\
      e_2 , \Sigma_2 \sreduce v , \Sigma_3}
      {\sif{e_1}{e_2}{e_3}, \Sigma_1 \sreduce v, \Sigma_3}
      
      \qquad

      \infer[RIfFalse]
      {e_1, \Sigma_1 \sreduce \code{false}, \Sigma_2 \\
        e_3 , \Sigma_2 \sreduce v , \Sigma_3}
      {\sif{e_1}{e_2}{e_3}, \Sigma_1 \sreduce v, \Sigma_3}
      
      \\ \\

      \infer[ROp]
      {e_1, \Sigma_1 \sreduce v_1, \Sigma_2 \\
       e_2, \Sigma_2 \sreduce v_2, \Sigma_3}
      {e_1 \oplus e_2, \Sigma_1 \sreduce v_1 \oplus v_2, \Sigma_3}

      \qquad

      \infer[RCstr]
      {e_i, \Sigma_i \sreduce v_i, \Sigma_{i+1} \\ i\in 1..n}
      {f(e_1, \ldots, e_n), \Sigma_1 \sreduce f(v_1, \ldots, v_n), \Sigma_{i+1}}

      \\ \\

      \infer[RInst]
      {
        e_1, \Sigma_1 \sreduce (\lambda x.e_2), \Sigma_2 \\
        H' = (\Sigma_2.H)[\sch \mapsto \lambda x.e_2]
      }
      {
        \sinstall \sch {e_1}, \Sigma_1 \sreduce \sunit, \Sigma_2[H
        \mapsto H']
      }

      \\ \\

      \infer[RProj]
      {e, \Sigma_1 \sreduce f(v_1, \ldots, v_n), \Sigma_2 }
      {f^{-i}(e), \Sigma_1 \sreduce v_i, \Sigma_2}

      \qquad 

      \infer[RSend]
      { e, \Sigma_1 \sreduce \xv, \Sigma_2 \\
        M' = (\Sigma_2.M) @ (\sch, \xv)
      }
      { \ssend \sch e, \Sigma_1 \sreduce \sunit, \Sigma_2[M \mapsto M']}
      \\ \\

      \\ \\ 

      \multicolumn{1}{l}{
        \framebox{$\Sigma_1 \treduce^{\evt} \Sigma_2$ \hbox{ and } 
          $\judge e \rightsquigarrow t$}
      }
      \\ \\

      \infer[THandle]
      { \Sigma_1.M = (\sch, \xv) @ M' \\
        \Sigma_1.H(\sch) = \lambda x.e \\\\
        \Sigma' = \Sigma_1[M\mapsto M'] \\
        e \aset{x \mapsto \xv} , \Sigma' \sreduce v, \Sigma_2 \\
      }
      { \Sigma_1 \treduce^{\tau} \Sigma_2}

      \qquad

      \infer[TInput]
      { \Sigma' = \Sigma[M \mapsto (\Sigma.M) @ (\sch, p)] }
      { \Sigma \treduce^{\sch!p} \Sigma' }

      \\ \\ 

      \infer[TOutput]
      { \Sigma.M = (\sfmt{netout},p), M' \\\\
        \Sigma' = \Sigma[M\mapsto M']
      }
      { \Sigma \treduce^{\sfmt{netout}!p} \Sigma'}

      \qquad

      

      \infer[TProg]
      {
      \Sigma_0 = ([(\code{onCreate},\code{unit})]
                  ,\emptyset
                  ,\aset{\code{onCreate} \mapsto
                          \lambda x . e})
      \quad x \not\in FV(e) 
        \\\\ 
      \Sigma_i \treduce^{\evt_i} \Sigma_{i+1}
      \quad i \in [0..n]
      }
      { \judge e \rightsquigarrow \evt_0 \cdot \evt_1 \cdots
      \evt_n }
    \end{array}
  \end{displaymath}
  \caption{Semantics for our Android subset.}
  \label{fig:semantics}
\end{figure*}

Our semantics is stratified into two levels: a big-step semantics,
shown at the top of Fig.~\ref{fig:semantics}, which models evaluation of code in a
handler, and a small-step semantics, shown at the bottom of the
figure, which models the message queue.  The semantics generates sets
of traces containing input and output messages.


The big-step semantics proves judgments of the form
$e, \Sigma_1 \sreduce v, \Sigma_2$, meaning
evaluation of expression $e$ in state $\Sigma_1$ produces a value $v$ and
new state $\Sigma_2$. 
The first several rules are standard.
\textsc{RVal} evaluates a value to itself, without changing the
state. \textsc{RApp} evaluates $e_1$ to a lambda, evaluates $e_2$ to
a value, and then evaluates the body of the lambda with the actual
argument substituted for the formal variable: we use the notation
$e\aset{x\mapsto v}$ for $e$ when unbound occurrences of $x$ in $e$
have been syntactically replaced with $v$. 
\textsc{RRef} evaluates $e$ to a value $v$, finds a fresh location
$\loc$ in the heap, and then evaluates to $\loc$, returning a state
where the heap maps $\loc$ to $v$. \textsc{RAssign} evaluates $e_1$
to a location $\loc$ and then updates the contents of
$\loc$. \textsc{RDeref} evaluates $e$ to a location and returns the
contents of that location.

\textsc{RIfTrue} evaluates $e_1$ to a value and if it evaluates to
\code{true}, evaluates $e_2$. \textsc{RIfFalse} is
analogous.
\textsc{ROp} evaluates a binary operation by applying the designated
operation to the values of the two subexpressions.

\textsc{RCstr} and \textsc{RProj} construct terms and project from
constructed terms. Finally, \textsc{RInst} evaluates $e_1$ to a lambda
and adds it as a handler for channel $\sch$; the result value is the
unit value $\sunit$ (a nullary constructor). \textsc{RSend} evaluates
$e$ to a primitive $\xv$, and then adds the message $(\sch,\xv)$ to the
end of the message queue. Here we use $@$ for concatenation. Note that \textsc{RSend} only allows
primitives to be sent, and not locations or lambda
expressions.

The first three rules 
in small-step semantics at the bottom of
Fig.~\ref{fig:semantics} prove judgments of the form
$\Sigma_1 \treduce^{\evt} \Sigma_2$, meaning the machine can take a step from state
$\Sigma_1$ to a new state $\Sigma_2$, producing an event $\evt$.
\textsc{THandle} consumes a message $(\sch, \xv)$ from the front of
the message queue (recall \textsc{RSend} adds a message to the
\emph{end}  of the message queue), looks up the handler for $\sch$,
and then invokes the handler, passing $\xv$ as its argument. Running a
handler is not an externally visible operation, which is indicated by an empty
event $\tau$ on the reduction arrow. 

\textsc{TInput} models an input message, which may be due to user
input (e.g., GUI clicks) or secret input from the system (e.g.,
a callback with updated location information). This rule
non-deterministically picks some channel $\sch$ and an arbitrary
primitive $p$, and then sends $p$ on that channel. 
The input is recorded as an event on the reduction arrow.
Notice that here we do not distinguish the security level of an input---we choose to leave that
up to the security policy designer, who can opt to either always
designate GUI inputs as low-security (as we do in our experiments) or
make them high-security.

\textsc{TOutput} models writes to the network, consuming a message
from a distinguished channel \sfmt{netout}. Writing to this channel
corresponds to the \code{InfoSender.sendInt} calls in
Section~\ref{sec:overview}. Since these messages may be seen by the
observer---i.e., they are ``low visible'' outputs---we record the
write event on the reduction arrow. Note that by convention
there is no user handler for this channel.

Finally, \textsc{TProg} proves a judgment of the form
$\judge e \rightsquigarrow t$, meaning running the program $e$ produces a
trace $t$ of events.
This rule creates an initial state $\Sigma_0$ in
which $e$ is bound as a handler on a special \code{onCreate} channel,
and the message queue contains an initial message on that channel. The
rule then repeatedly steps to the next state $n+1$ times. It produces 
the event trace $\eta_0\cdots\eta_n$.
Notice that the length of the trace $n$ is nondeterministic; in general,
since these are reactive programs, they can usually run for an any
number of steps as long as additional input arrives.

From the set of all executions of a program we can extract a set of
traces, which we later use to define noninterference.
\begin{definition}[Program Traces]
  We define the set of traces of a program $e$ as
\begin{displaymath}
  \traces(e) = \aset{t \mid \; \judge e { \rightsquigarrow t }}
\end{displaymath}
\end{definition}

\section{Linear Temporal Logic}
\label{sec:ltl}

Figure~\ref{fig:ltl} gives the definition of the models relaion for
LTL as used in \ref{sec:formalism}.  Note that we assume an
interpretation $\models p_1 \oplus p_2$ for the atomic propositions of
the form $p_1 \oplus p_2$.

\begin{figure}[!t]
  \small
  \begin{displaymath}
    \begin{array}{rcl}
      \tr , i \models \sch {!}p_1 & \iff & \tr[i] = \sch {!}p_1 \\
      \tr , i \models \sch {!}* & \iff & \hbox{there exists } p,
      \tr[i] = \sch {!}p_1 \\
      \tr , i \models p_1 \oplus p_2 & \iff & \models p_1 \oplus p_2 \\
      
      \tr, i \models \neg \phi & \iff &
      \tr, i  \not \models \phi \\

      \tr, i  \models \phi \land \psi & \iff &
      \tr, i  \models \phi \hbox{ and } \tr, i  \models \psi \\

      \tr, i  \models \phi \lor \psi & \iff &
      \tr, i  \models \phi \hbox{ or } \tr, i  \models \psi \\

      \tr, i  \models \phi \limplies \psi & \iff &
      \tr, i  \models \psi \hbox{ or } \tr, i  \models
      \lnot \phi \\

      \tr, i  \models \talways \phi & \iff &
      \hbox{for all } j ~.~ j \geq i \mimplies (\tr, j  \models \phi) \\

      \tr, i  \models \tfuture \phi & \iff &
      \hbox{there exists } j, j \geq i \mimplies (\tr, j  \models \phi) \\

      \tr, i  \models \tpast \phi & \iff &
      \hbox{there exists } j, j \leq i \mimplies (\tr, j  \models \phi) \\

      \tr, i  \models \phi \tuntil \psi & \iff &
      \hbox{there exists } j, j \geq i \mimplies ((\tr, j  \models \psi)
      \hbox{ and } \\
      & & \hbox{for all } k, i \leq k < j \mimplies (\tr, k 
      \models \phi)) \\

      \tr, i  \models \phi \tsince \psi & \iff &
      \hbox{there exists } j, j \leq i \mimplies ((\tr, j  \models \psi)
      \hbox{ and } \\
      & & \hbox{for all } k, j < k \leq i \mimplies (\tr, k 
      \models \phi)) \\

      \tr, i  \models \forall x. \phi & \iff &
      \hbox{for all } p, (\tr, i  \models \aset{x\mapsto p} \phi) \\

      \tr, i  \models \exists x. \phi & \iff &
      \hbox{there exists } p, (\tr, i  \models \aset{x\mapsto p} \phi) \\
    \end{array}
  \end{displaymath}
  \caption{Models relation for LTL used in the paper}
  \label{fig:ltl}
\end{figure}

\end{document}